\documentclass{appolb}
\usepackage{graphicx}

\preprint{}

\begin{document}

\title{{\bf Thermal approach to RHIC}\thanks{Lecture presented  by WB at the 
{\em XLII Cracow School of Theoretical Physics}, Zakopane, Poland, 31 May -- 9 June 2002}
\thanks{Research supported in part by the Polish State Committee for
Scientific Research, grant number 2~P03B~09419.}}
\author{Wojciech Broniowski, Anna Baran, \\ Wojciech Florkowski 
\address{The H. Niewodnicza\'nski Institute of Nuclear Physics\\
ul. Radzikowskiego 152, PL-31342 Krak\'ow, Poland}}

\maketitle

\begin{abstract}
Applications of a simple thermal model to ultra-relativistic heavy-ion
collisions are presented.
We compute abundances of various hadrons, including particles 
with strange quarks, the $p_\perp$-spectra, and the HBT radii for the pion. 
Surprising agreement is found, showing that the 
thermal approach can be used successfully to understand and describe the RHIC data.
\end{abstract}


\thispagestyle{empty}

\section{Introduction\label{sec:intro}}

With the wide stream of new high-quality data flowing from RHIC, as
well as with the continued efforts at SPS (for recent results see,
{\em e.g.}, \cite{qm01,qm02,hir02}), there is a growing need for a
simple description of the basic underlying physics. Only then our
understanding of the phenomena occurring in ultra-relativistic
heavy-ion collisions can be pushed forward, and space made for
potential new phenomena, hitherto unexplained within the basic
picture. In this lecture we argue that most of the ``soft'' features
of the data from RHIC (particle ratios, momentum spectra, HBT
correlation radii) can be explained very efficiently within an
embarrassingly simple model, which merges the thermal model
\cite{braunmu,raf,cest,pbmsps,yg,becatt,gazgor0,gaz,raf0,pbmrhic,wfwbmm,mm,budzan,gorj,bps,bg,cl,z} with
expansion, and incorporates in a complete way the resonances
\cite{wbwf,wfepi,hirsch,str,rhicsps,wfqm02}. Our description uses
hadronic degrees of freedom and starts at freeze-out, {\em i.e.}  at
the point of the space-time evolution of the system where the hadrons
cease to interact. Pertinent theoretical questions, such as what had
been happening before freeze-out, what had led to the strong expansion
of the system, why is the thermal picture successful, what is the
nature of hadronization, not to mention the notorious {\em ``was there
quark-gluon plasma?''}, will not and cannot be addressed in this
lecture. Nevertheless, we believe that our studies prepare ground for
such questions.

\section{The thermal model\label{sec:model}}

Historically, the ideas of the thermal description of a hadronic
system go back to the works of Koppe \cite{koppe}, Fermi \cite{fermi},
Landau \cite{landau}, and Hagedorn \cite{hagedorn}.  More recently,
many groups have used these ideas in numerous papers in an effort to
explain the data from various relativistic heavy-ion experiments, from
SIS, through AGS and SPS, to RHIC. Along the way, the original picture
has been occasionally supplied with extra features, such as the
fugacities controlling deviations from chemical equilibrium
\cite{gammas}, finite volume and Van der Waals corrections
\cite{braunmu,vdw}, or the use of the canonical instead of the
grand-canonical ensemble \cite{rafdan,hamieh,rafcrit}.

The works of Heinz and collaborators \cite{heinzr} put forward the
concept of two freeze-outs. As the system expands and cools, it first
passes through the chemical freeze-out point at temperature $T_{\rm
chem}$. Later, the particles can only rescatter elastically, until
these processes are switched off at a lower temperature $T_{\rm
kin}$. In an appealing way the distinct freeze-outs explained the need
for a higher temperature to reproduce well the particle ratios, and a
much lower temperature to describe the slopes of the momentum spectra.
In our work \cite{wfwbmm,wbwf,str} we have shown that with the
complete treatment of resonances, the distinction between the two
freeze-outs is not needed, at least for RHIC, and one can achieve very
good explanation of all ``soft'' features of data assuming one
universal freeze-out,
\begin{equation}
T_{\rm chem}=T_{\rm kin} \equiv T. \label{Tuniv}
\end{equation}
We have also dropped, with the Ockham razor at hand, all other
additions to the most naive thermal approach.  The dropped features
may be reconsidered later on, provided there is a well-established
phenomenological need, or theoretical argumentation.

\begin{figure}[b]
\centerline{\includegraphics[width=9.5cm]{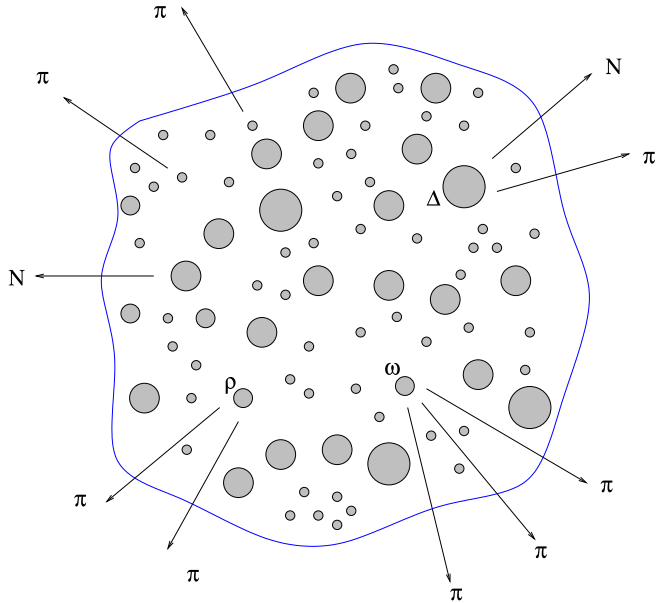}}
\vspace{-2.3cm}
\caption{A schematic picture of the hadronic soup, 
formed in an ultra-relativistic heavy-ion collision at freeze-out. 
The resonances decay subsequently into stable particles. The inclusion of 
many resonances is crucial for the success of the thermal approach.}
\label{fig:therm}
\end{figure}

The main ingredients of our model are as follows:

\begin{itemize} 

\item
There is one freeze-out, as discussed above, at which all the hadrons
occupy the available phase space according to the statistical
distribution factors.  The scenario with a single freeze-out is
natural if the hadronization occurs in such conditions that neither
elastic nor inelastic processes are effective. An example here is the
picture of the supercooled plasma of Ref. \cite{rafelski}.  Moreover,
the STAR collaboration \cite{starKstar,starKstar2} has presented an important
argument in favor of very weak rescattering after the chemical
freeze-out at RHIC, based on the observation of the $K^*(892)^0$ peaks
in the pion-kaon correlations.  This essentially shows that the
expansion time between the chemical and thermal freeze-out is shorter
than the life-time of the $K^*(892)^0$, {\em i.e.} $\sim 4$
fm/$c$. Additionally, the fact that the measured yields of
$K^*(892)^0$ \cite{starKstar,starKstar2} are reproduced very well within the
thermal model \cite{pbmrhic,wfwbmm} hints to the scenario with a short
time between the two possible freeze-outs, as proposed in
Ref. \cite{wbwf}. Thus, approximation (\ref{Tuniv}) is reasonable.

\item
A crucial feature of our analysis is the {\em complete} treatment of
the hadronic states, with all resonances from the Particle Data Table
\cite{PDG} included in the analysis of both the ratios and the
momentum spectra (cf. Fig. \ref{fig:therm}).  Although the high-lying
states are suppressed by the thermal factors, their number increases
according to the Hagedorn hypothesis
\cite{hagedorn,myhag,bled,rafhag}, such that their net effect is
important. For instance, only a quarter of the observed pions at RHIC
comes from the ``primordial'' pions present at freeze-out, and three
quarters are produced via resonance decays. All decays, two and three
body, are implemented in cascades, with the branching ratios taken
from the tables. For the $p_{\perp}$-spectra the resonances are also
very important, since their decays increase the slope, as if the
temperature were effectively lower \cite{wfwbmm}. It has also been found 
that the inclusion of resonances speeds up the cooling of the system 
in hydrodynamic calculations \cite{hirano}.

\item Whereas expansion of the system does not alter the particle
ratios at midrapidity (provided the system is boost-invariant, which
is a good approximation at RHIC, see Sec. \ref{sec:expansion}), it
becomes absolutely essential for the $p_{\perp}$-spectra. We model the
expansion (transverse flow) and the size of the system with two
parameters: the proper time at freeze-out, $\tau$, and the transverse
size, $\rho_{\rm max}$.

\item The model has altogether four adjustable parameters: two thermal
and two geometric, which possess clear physical interpretation. The
two thermal parameters, temperature, $T$, and the baryon chemical
potential, $\mu_B$, are fixed by the analysis of the ratios of the
particle abundances \cite{wfwbmm}.  The two geometric parameters are
fixed with help of the $p_\perp$-spectra.  The invariant time $\tau$
controls the
overall normalization of the spectra, while the ratio $\rho_{\rm
max}/\tau$ directly influences their slopes.

\end{itemize}

All data used in the present study are for Au+Au collisions at
$\sqrt{s_{NN}}=130$~GeV.  

\section{Particle ratios\label{sec:ratios}}

\begin{table}[t]
\begin{centering}
\begin{tabular}{|l|r|c|}
\hline
& Model & Experiment \\ \hline \hline
\multicolumn{3}{|l|}{Fitted thermal parameters}\\ \hline\hline
$T$ [MeV] & 165$\pm 7$ &  \\ \hline
$\mu _{B}$ [MeV] & \ 41$\pm 5$ &  \\ \hline
$\mu _{S}$ [MeV] & \ \ \ \ \ 9 &  \\ \hline
$\mu _{I}$ [MeV] & \ \ \ \ \ -1 &  \\ \hline
$\chi ^{2}/n$ & 0.97 &  \\ \hline \hline
\multicolumn{3}{|l|}{Ratios used for the fit}\\ \hline\hline
$\pi ^{-}/\pi ^{+}$ & $1.02$ & 
\begin{tabular}{ll}
$1.00\pm 0.02$ \cite{phobos}, & $0.99\pm 0.02$\cite{bearden}
\end{tabular}
\\ \hline 
$\overline{p}/\pi ^{-}$ & $0.09$ & $0.08\pm 0.01$ \cite{harris} \\ \hline
$K^{-}/K^{+}$ & $0.92$ & 
\begin{tabular}{ll}
$0.88\pm 0.05$ \cite{caines}, & $0.78\pm 0.12$ \cite{ohnishi} \\ 
$0.91\pm 0.09$ \cite{phobos}, & $0.92\pm 0.06$ \cite{bearden}
\end{tabular}
\\ \hline
$K^{-}/\pi ^{-}$ & $0.16$ & $0.15\pm 0.02$ \cite{caines} \\ \hline
$K_{0}^{\ast }/h^{-}$ & $0.046$ & $0.060\pm 0.012$ \cite{caines,zxu} \\ 
      &  & later: $0.042 \pm 0.011$ \cite{starKstar2} \\ \hline
$\overline{K_{0}^{\ast }}/h^{-}$ & $0.041$ & $0.058\pm 0.012$ \cite{caines,zxu} \\ 
      &  & later: $0.039 \pm 0.011$ \cite{starKstar2} \\ \hline
$\overline{p}/p$ & $0.65$ & 
\begin{tabular}{ll}
$0.61\pm 0.07$ \cite{harris}, & $0.54\pm 0.08$ \cite{ohnishi} \\ 
$0.60\pm 0.07$ \cite{phobos}, & $0.61\pm 0.06$ \cite{bearden}
\end{tabular}
\\ \hline
$\overline{\Lambda }/\Lambda $ & $0.69$ & $0.73\pm 0.03$ \cite{caines} \\ 
\hline
$\overline{\Xi }/\Xi $ & $0.76$ & $0.82\pm 0.08$ \cite{caines} \\ \hline \hline
\multicolumn{3}{|l|}{Ratios predicted}\\ \hline\hline
$\phi/h^-$ & $0.019$ & $0.021 \pm 0.001$ \cite{starphi} \\ \hline
$\phi/K^-$ & $0.15$ & 0.1 -- 0.16 \cite{starphi} \\ \hline
$\Lambda/p$ & 0.47 & $0.49 \pm 0.03$ \cite{starLambda,starantip} \\ \hline
$\Omega^-/h^-$ & 0.0010 & $0.0012 \pm 0.0005$ \cite{suire} \\ \hline
$\Xi^-/\pi^-$ & 0.0072 & $0.0085 \pm 0.0020$ \cite{castilloxi} \\ \hline
$\Omega^+/\Omega^-$ & 0.85 & $0.95 \pm 0.15$ \cite{suire} \\ \hline
\hline
\end{tabular}
\caption{Optimal thermal parameters, ratios $\left
. \frac{dN_i/dy}{dN_j/dy} \right | _{y=0}$ used for the fit, and
further predicted ratios. The preliminary experimental numbers for
$K^\ast(892)$ \cite{zxu} have changed \cite{starKstar2}, and better agreement with the
model followed.}
\end{centering}
\label{tab:fit}
\end{table}

The density of the $i$th hadron species is calculated 
from the ideal-gas expression
\begin{eqnarray}
n_{i}&=&g_{i} \int d^3p \, f_{i}(p), \nonumber \\
f_{i}(p)&=&\frac{1}{(2\pi)^3} \left ( {\exp \left[
\left( E_{i}(p)-\mu _{B}B_{i}-\mu _{S}S_{i}-\mu
_{I}I_{i}\right) /T\right] \pm 1} \right )^{-1},  \label{ni}
\end{eqnarray}
where $g_{i}$ is the spin degeneracy, $ B_{i}$, $S_{i}$, and $I_{i}$
denote the baryon number, strangeness, and the third component of
isospin, and $E_{i}(p)=\sqrt{p^{2}+m_{i}^{2}}$. The quantities $ \mu
_{B}$, $\mu _{S}$, and $\mu _{I}$ are the chemical potentials enforcing
the appropriate conservation laws. We recall that Eq. (\ref{ni}) is
used to calculate the ``primordial'' densities of stable hadrons as
well as of all resonances at the freeze-out, which later on decay.
The temperature, $T$, and the baryonic chemical potential, $\mu_{B}$,
have been fitted with the $\chi^2$ method to the originally available
experimental ratios of particles, listed in the second group of rows
in Table 1. The $\mu_{S}$ and $\mu_{I}$ are determined with the
conditions that the initial strangeness of the system is zero, and the
ratio of the baryon number to the electric charge is the same as in
the colliding nuclei. It turns out that the role of $\mu_{I}$ at RHIC
is negligible.
 
For boost-invariant systems the ratios of hadron multiplicities at
midrapidity, $dN/dy|_{y=0}$, are related to the ratios of densities,
$n_i$, since
\begin{equation}
\left . \frac{dN_i/dy}{dN_j/dy} \right |_{y=0}=\frac{N_i}{N_j}=\frac{n_i}{n_j}. 
\label{dN}
\end{equation}
The first equality follows trivially from the 
assumed boost invariance, while
second one reflects the factorization of the volume of the system (%
see Sec.~\ref{sec:expansion}). Hence the midrapidity ratios,
$\left . \frac{dN_i/dy}{dN_j/dy} \right | _{y=0}$, may be used 
to fit the thermal parameters of the model.

Table 1 presents the result of the fit. In our procedure the ratios measured by
different groups enter separately in the definition of
$\chi^{2}$. Thus, the number of the used data points is $n=16$. The
obtained optimal value of $T=165\pm 7$~MeV is, most interestingly,
consistent with the value of the critical temperature for the
deconfinement phase transition obtained from the QCD lattice
simulations: $T_{c}=154\pm 8$~MeV for three massless flavors and
$T_{c}=173\pm 8$~MeV for two massless flavors \cite{Karsch}.  We note
that our $T$ is 9 MeV lower than 174~MeV of Ref. \cite{pbmrhic}, and
25 MeV lower than 190~MeV obtained in Ref. \cite{nxu}. Nevertheless,
the results of the three calculations are consistent within errors. We
have also computed other characteristics of the freeze-out: the energy
density, $\varepsilon =0.5$ GeV/fm$^{3}$, the pressure, $P=$ 0.08
GeV/fm$^{3}$, and the baryon density, $\rho _{B}=$ 0.02 fm$^{-3}$.  We
note that the results for the $K^\ast(892)$ mesons, off by 50\% when
compared to the early preliminary data \cite{zxu}, came within the
error bars of the data corrected later \cite{starKstar,starKstar2}.
The lower part of Table \ref{tab:fit} contains our predictions for
particles containing strange quarks.  The agreement with the data,
released later, is very good. In particular, the triply-strange 
$\Omega$ is properly reproduced.

To summarize this part, we stress the high quality of the fit in Table
1 for all kinds of particles, including those carrying strange quarks.

\section{Expansion\label{sec:expansion}}

Obviously, much richer information on the hadron production is
contained in the transverse-momentum spectra. Various collaborations
at RHIC measure, with impressive accuracy, the particle spectra of
different hadrons, \mbox{$dN_i/(2\pi p_\perp dp_\perp dy)$}, at
midrapidity and for various centrality bins (the latter may be mapped
to different impact parameters, \cite{centr}). Unlike the case of the
ratios of Sec. \ref{sec:ratios}, modeling of the spectra involves not
only setting the thermal parameters, but also a suitable inclusion of
the expansion. Clearly, hydrodynamic flow modifies the spectra via the
Doppler effect.  Thus, an important ingredient of our model is the
choice of the freeze-out hypersurface ({\em i.e.} a three-dimensional
volume in the four-dimensional space-time) and the four-velocity field at
freeze-out.  Many choices are possible here, with some hinted by the
hydrodynamic calculations. Our choice has been made in the spirit of
Ref.~\cite{bjorken,baym,Kolya,siemens,SSH,BL,Rischke,SH}, and is
defined by the condition
\begin{equation}
\tau = \sqrt{t^2-r^2_x-r^2_y-r^2_z} = {\rm const}.
\label{tau}
\end{equation}
Later on we denote the constant in Eq. (\ref{tau}) simply by $\tau$.
In order to make the transverse size, 
\begin{equation}
\rho=\sqrt{r_x^2+r_y^2}, \label{rhodef}
\end{equation}
finite, we impose the condition $\rho < \rho_{\rm max}$. In addition,
we assume that the four-velocity of the hydrodynamic expansion at
freeze-out is proportional to the coordinate (Hubble-like expansion),
\begin{equation}
u^{\mu } =\frac{x^{\mu }}{\tau }=\frac{t}{\tau }\left(
1,\frac{r_{x}}{t},\frac{r_{y}}{t},\frac{r_{z}}{t}\right).
\label{umu}
\end{equation}
Such a form of the flow at freeze-out, as well as the fact that $t$
and $r_z$ coordinates are not limited and appear in the
boost-invariant combination in Eq. (\ref{tau}), imply that our
model is boost-invariant. We have checked numerically that this
approximation works very well for calculations in the central-rapidity
region.

In practical calculations it is convenient to introduce the
following parameterization \cite{BL}:
\begin{eqnarray}
t &=&\tau \cosh \alpha _{\parallel }\cosh \alpha _{\perp },\quad r_{z}=\tau
\sinh \alpha _{\parallel }\cosh \alpha _{\perp },  \nonumber \\
r_{x} &=&\tau \sinh \alpha _{\perp }\cos \phi ,\quad r_{y}=\tau \sinh \alpha
_{\perp }\sin \phi ,  \label{par}
\end{eqnarray}
where $\alpha _{\parallel }$ is the rapidity of the fluid element,
$v_{z}=r_z/t=\tanh \alpha _{\parallel }$, and $\alpha _{\perp }$ describes
the transverse size, $\rho =\tau \sinh \alpha _{\perp }$.
The transverse velocity is 
$v_\rho=\tanh \alpha_\perp/\cosh \alpha_\parallel$. 
The element of the hypersurface is defined as
\begin{equation}
d\Sigma_\mu
= \epsilon_{\mu \alpha \beta \gamma}
{\partial x^\alpha \over \partial \alpha_\parallel}
{\partial x^\beta \over \partial \alpha_\perp}
{\partial x^\gamma \over \partial \phi} \, d\alpha_\parallel 
d\alpha_\perp d\phi, 
\label{sigma}
\end{equation}
where $x^0=t$, $x^1=r_x$, $x^2=r_y$, $x^3=r_z$ and $\epsilon_{\mu
\alpha \beta \gamma}$ is the Levi-Civita tensor. A straightforward
calculation yields
\begin{equation}
d\Sigma^\mu(x) = u^\mu(x)\, \tau ^{3} \, {\rm sinh}(\alpha _{\perp})
{\rm cosh}(\alpha _{\perp}) \, d\alpha _{\perp}
d\alpha _{\parallel } d\phi,
\label{prop}
\end{equation}
such that the four-vectors $d\Sigma^\mu$ and $u^\mu$ turn out to be
parallel. This feature is special for our choice (\ref{tau},\ref{umu}),
and in general does not hold.

A question comes to mind as to what extent the assumptions
(\ref{tau},\ref{umu}) are realistic from the point of view of
hydrodynamics.  As a results of a typical hydrodynamic calculation,
the freeze-out hypersurface contains, in the $\rho$-$t$ plane, a
time-like and a space-like parts
\cite{baym,Kolya,siemens,SSH,BL,Rischke,SH}. The latter one is plagued
with conceptual problems \cite{bugaev,csernai,neymann,magas}. Our
parameterization neglects the space-like part altogether, thus avoiding
difficulties.  The time-like part of the hypersurface has, in many
hydrodynamic calculations, the feature that the outer regions in the
transverse direction freeze out earlier than the inner regions. Our
choice (\ref{tau}), as well as commonly used versions of the
blast-wave model, where the freeze-out occurs at a constant value of
$t$, do not share this feature.  On the contrary, our
Eqs. (\ref{tau},\ref{umu}) correspond to the so-called {\em scaling
solution} \cite{baym,CF2,biro} of hydrodynamic equations, which is
obtained in the case where the sound velocity in the medium is low.
Naturally, the validity of the assumptions and their relevance for the
results should be examined in a greater detail. In Ref. \cite{wbwf} we
have checked that two different models of the expansion lead to very
close predictions for the momentum spectra at RHIC. Other
parameterizations may be also verified with the help of the formulas
given below. The fact that parameterization (\ref{tau},\ref{umu})
works impressively well (cf. Sec. \ref{sec:spectra}), and at the same
time the conventional hydrodynamic calculations have serious
problems in explaining the RHIC data, hints, in our opinion, for a
revision of the part of the assumptions entering hydrodynamic
calculations and for extensions \cite{finland,csorgohyd,csster} of the picture
used up to now.

\section{Decays of resonances \label{sec:decofres}}

The decays of resonances present a technical complication in the
formalism. The resonances are formed on the freeze-out hypersurface
with a given four-velocity. In the local rest frame of the fluid
element the momenta of the resonances have thermal distribution,
however, their decay products have, obviously, a different
(non-thermal) distribution, which reflects the 
distribution of the resonance and the kinematics.  
Below, we describe in detail our method,
which is exact and semi-analytic (final expressions involve simple
numerical integration rather than involved Monte-Carlo simulations).

\begin{figure}[t]
\centerline{\includegraphics[width=9.5cm]{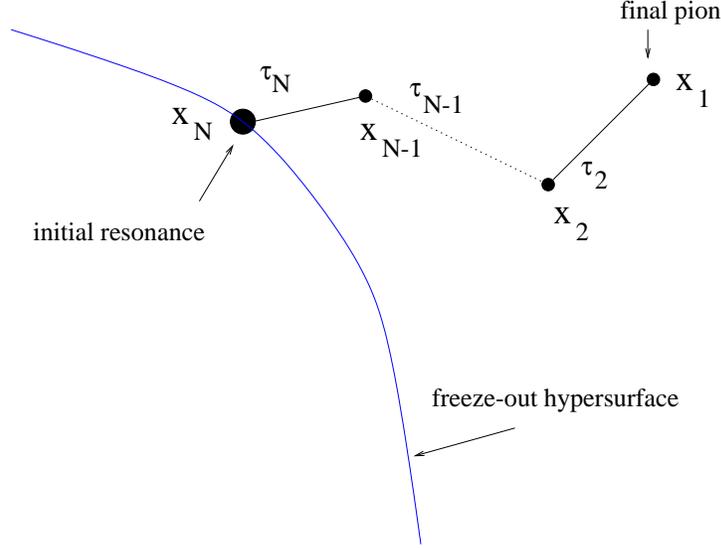}}
\vspace{0cm}
\caption{The cascade of resonance decays.}
\label{fig:resdec}
\end{figure}

Consider a sequence of the resonance decays of
Fig. \ref{fig:resdec}. The initial resonance decouples on the
freeze-out hypersurface at the space-time coordinate $x_N$, and
decays after time $\tau_N$, with an average time proportional to the
life-time $1/\Gamma_N$.
\footnote{In this section the indices $i$ label the position in the cascade, and not 
the particle species, as in Sec. \ref{sec:expansion}.}
Let us track a single decay product.  It is
formed at the point $x_{N-1}$, decays again after time $\tau_{N-1}$,
and so on.  At the end of the cascade a particle with label 1 is
formed, which is being detected.  The Lorentz-invariant phase-space
density of the measured particles is
\begin{eqnarray}
&&n_{1 }\left( x_{1},p_{1}\right)  =  \label{npix1p1} \\ 
&& \int \frac{d^{3}p_{2}}{E_{p_{2}}}
B\left( p_{2},p_{1}\right) \int d\tau _{2}\Gamma _{2}e^{-\Gamma _{2}\tau
_{2}} \int d^{4}x_{2}\delta ^{\left( 4\right) }
\left( x_{2}+\frac{p_{2}\tau_{2}}{m_{2}}-x_{1}\right)...  \nonumber \\
&&\times \int \frac{d^{3}p_{N}}{E_{p_{N}}}B\left( p_{N},p_{N-1}\right) \int
d\tau _{N}\Gamma _{N}e^{-\Gamma _{N}\tau _{N}} \nonumber \\
&&   \hspace{5mm} \times \int d\Sigma _{\mu }\left(
x_{N}\right) \,p_{N}^{\mu }\,\,\delta ^{\left( 4\right) }\left( x_{N}+\frac{%
p_{N}\,\tau _{N}}{m_{N}}-x_{N-1}\right) f_{N}\left[ p_{N}\cdot u\left(x_{N}\right) 
\right]. \nonumber
\end{eqnarray}
We have generalized here the formula from Ref. \cite{ornik} where a single
resonance decay, without cascades, is taken into account.  Note that
the integration over $x_{N-1}\, ... \,\, x_2$ is unconstrained, while
the integration over $x_N$ is constrained to the hypersurface
$\Sigma$. The delta functions impose the condition that the particle
of velocity $p_n/m_n$ travels the distance from $x_n$ to $x_{n-1}$ in
time $\tau_n$.  The function $B(q,k)$ is the probability distribution for a
resonance with momentum $k$ to produce a particle with momentum $q$,
namely
\begin{equation}
B(q,k) = {b \over 4 \pi p^*} \delta \left( {k \cdot q \over m_R}- E^* \right),
\label{B}
\end{equation} 
where $b$ denotes the branching ratio for the particular decay
channel,
\footnote{In this notation $b$ includes also the ratio
of the spin degeneracies of the two particles.} and $p^*$ ($E^*$) is
the momentum (energy) of the emitted particle in the resonance's rest
frame.  We note that $B(k,q)$ satisfies the normalization condition
\begin{equation}
\int {d^3q \over E_q} B(q,k) = b.
\label{normB}
\end{equation}
Integration over all space-time positions in Eq. (\ref{npix1p1}) gives
the formula for the momentum distribution
\begin{eqnarray}
&& E_{p_1} {dN_1 \over d^3 p_1} = \int d^{4}x_{1}\,n_{1}
\left(x_{1},p_{1}\right) = \label{npip1} \\
&&\int \frac{d^{3}p_{2}}{E_{p_{2}}}B\left( p_{2},p_{1}\right)...\int 
\frac{d^{3}p_{N}}{E_{p_{N}}}B\left( p_{N},p_{N-1}\right) 
\int d\Sigma_{\mu }\left( x_{N}\right) \,p_{N}^{\mu }\,\,f_{N}
\left[ p_{N}\cdot u\left(x_{N}\right) \right], \nonumber
\end{eqnarray}
which should be used in the general case of any $\Sigma$ and $u$.
\footnote{Note that the dependence on the widths $\Gamma_k$ has
disappeared, reflecting the fact that for the momentum spectra it is
not relevant when or where the resonances decay. It is important,
however, when and where the particles {\it decouple} from each
other, which is determined by the choice of $\Sigma(x_N)$ and $u(x_N)$.} 

We are now going to prove the second equality in Eq. (\ref{dN}). 
Starting from Eq.  (\ref{npix1p1}) we find the multiplicity
of particles of type 1 coming from the discussed chain decay,
\begin{equation}
N_1  = b_{N\rightarrow N-1} \, ... \, b_{2\rightarrow 1} 
\int d\Sigma_{\mu }\left( x_{N}\right) \int
\frac{d^{3}p_{N}}{E_{p_{N}}} \,p_{N}^{\mu }\,\,f_{N}
\left[ p_{N}\cdot u\left(x_{N}\right) \right], 
\label{mulN1}
\end{equation}
with an obvious notation for the branching ratios.
Since the last integral in Eq. (\ref{mulN1}) yields an expression proportional to $u^\mu(x_N)$, 
and the distribution function of the resonance $N$ is thermal, we can
rewrite Eq. (\ref{mulN1}) in the equivalent form
\begin{eqnarray}
N_1  &=& b_{N\rightarrow N-1}
 \, ... \, b_{2\rightarrow 1} \int d\Sigma_{\mu }\left( x_{N}\right) u^\mu(x_N) 
n_N\left[T(x_N),\mu_B(x_N) \right] \nonumber \\
&=& b_{N\rightarrow N-1}
 \, ... \, b_{2\rightarrow 1} 
n_N\left(T,\mu_B\right)
\int d\Sigma_{\mu }\left( x_{N}\right) u^\mu(x_N) .
\label{mulN2}
\end{eqnarray}
Eq. (\ref{mulN2}) indicates that the volume factor at
freeze-out, $\int d\Sigma_{\mu }\left( x_{N}\right) u^\mu(x_N)$, 
factorizes if the thermodynamic conditions (temperature and
chemical potentials) are constant on the freeze-out hypersurface. This
observation leads directly to the general conclusion that, as long as
we integrate (measure) the spectra in the full phase-space (or, for boost-invariant 
systems, at a given rapidity $y$), the ratios
of the particle yields are not affected by the flow and can be
calculated with help of the simple expressions valid for static
systems. This completes the proof of the second equality
in Eq. (\ref{dN}). 

An important simplification follows if the element of the freeze-out
hypersurface is proportional to the four-velocity. This is precisely
the case considered in our model where (compare Eq. (\ref{prop}))
\begin{equation}
d\Sigma_{\mu }(x_{N})=d\Sigma(x_{N}) \,u_\mu(x_{N}).
\label{A4}
\end{equation}
Then
\begin{eqnarray}
E_{p_1} {dN_1 \over d^3 p_1}&=&\int d\Sigma \left(
x_{N}\right) \int \frac{d^{3}p_{2}}{E_{p_{2}}}
B\left( p_{2},p_{1}\right) ... \nonumber \\
&& \times
\int \frac{d^{3}p_{N}}{E_{p_{N}}}B\left( p_{N},p_{N-1}\right) p_{N}
\,\cdot u\left( x_{N}\right) \,f_{N}\left[ p_{N}\cdot u\left( x_{N}\right)
\right]  \nonumber \\
&=&\int d\Sigma \left( x_{N}\right)
p_{1}\,\cdot u\left( x_{N}\right)
\,f_{1}\left[ p_{1}\cdot u\left( x_{N}\right) \right],
\label{A5}
\end{eqnarray}
where we have introduced
\begin{eqnarray}
& & p_{i-1}\,\cdot u\left( x_{N}\right) \,f_{i-1}\left[ p_{i-1}\cdot
u\left( x_{N}\right) \right] \label{trsim} \\& & =\int \frac{d^{3}p_{i}}{E_{p_{i}}}B\left(
p_{i},p_{i-1}\right) p_{i}\,\cdot u\left( x_{N}\right) \,f_{i}\left[
p_{i}\cdot u\left( x_{N}\right) \right]. \nonumber 
\end{eqnarray}
The meaning of Eq. (\ref{trsim}) is that as we step down along the
cascade, the momentum distribution of the decay product, $f_{i-1}$, is
obtained from the momentum distribution of the decaying particle,
$f_i$, with a simple integral transform following from the kinematics.
In the fluid local-rest-frame, most convenient in the numerical
calculation, we have $u^\mu(x_N)=(1,0,0,0)$, and the transformation
(\ref{trsim}) reduces to the form \cite{wfwbmm}
\begin{equation}
f_{i-1}\left( q\right) = 
\frac{b m_R}{2 E_q p^\ast q} \int_{k_{-}(q)}^{k_{+}(q)} dk\, k \,
f_i\left( k\right),  \label{ftildecom}
\end{equation}
where the limits of the integration are $k_{\pm }(q)=m_{R} \left|
E^\ast q \pm p^\ast E_q \right| / m_1^2 $.  Equation~(\ref{ftildecom}) is a
relativistic generalization of the expression derived in
Ref.~\cite{weinhold}. 
The technical advantage of Eq. (\ref{A5}) is that the cascade 
can be performed in the rest frame of the original particle, with 
spherical symmetry and
one-dimensional integrations over momenta, (\ref{ftildecom}), while in the 
general case of Eq. (\ref{npip1}) only cylindrical symmetry holds and
two-dimensional integrations over momenta must be used.

In the case of three-body decays we follow the same steps as above, with
a modification arising from the fact that now different values of $
p^{\ast }$ are kinematically possible. This introduces an additional integration in
Eq. (\ref{ftildecom}). The distribution of the allowed values of $p^{\ast}$ may
be obtained from the phase-space integral 
\begin{equation}
A\int \frac{d^{3}p_{1}}{E_{p1}}\frac{%
d^{3}p_{2}}{E_{p_{2}}}\frac{d^{3}p_{3}}{E_{p_{3}}}\delta \left(
m_{R}\!-\!E_{p_{1}}\!-\!E_{p_{2}}\!-\!E_{p_{3}}\right) \delta ^{(3)}\left( 
{\bf p}_{1}\!+\!{\bf p}_{2}\!+\!{\bf p}_{3}\right) \left| {\cal M}\right| ^{2}, 
\label{gp1}
\end{equation}
where ${\bf p}_{1},{\bf p}_{2}$ and ${\bf p}_{3}$ are the momenta of
the emitted particles, $E_{p_{1}},E_{p_{2}}$ and $E_{p_{3}}$ are the
corresponding energies (all measured in the rest frame
of the decaying particle), ${\cal M}$
is the matrix element describing the three-body decay, and $A$ is a
normalization factor.  For simplicity we assume, similarly
to \cite{SKH}, that ${\cal M}$ can be approximated by a
constant, {\em i.e.} only the phase-space effect is included. 
Operationally, the final expression for three-body decays is
a folding of two-body decays over $p^{\ast }$ with a weight following
from elementary considerations based on Eq. (\ref{gp1}).
 
Finally, for the case satisfying condition (\ref{A4}), the spectra are
obtained from the expression analogous to the Cooper-Frye \cite{CF2,CF1} formula,
\begin{equation}
\frac{dN}{d^{2}p_{\perp }dy} =
\int p^{\mu }d\Sigma _{\mu }\ f_{1}\left(p\cdot u\right) ,
\label{Ni}
\end{equation}
but with the distribution $f_1$ which has collected the products of
resonance decays. With parameterization (\ref{par}) we can rewrite
Eq. (\ref{Ni}) in the form
\begin{eqnarray}
\frac{dN}{d^{2}p_{\perp }dy} &=&\ \tau ^{3}\int_{-\infty }^{+\infty
}d\alpha _{\parallel }\int_{0}^{\rho _{\max }/\tau }{\rm sinh}  \alpha _{\perp
}d\left( {\rm sinh}  \alpha _{\perp }\right) \int_{0}^{2\pi }
d\xi \, p\cdot u \, f_{1}\left( p\cdot u\right) ,\nonumber \\
\label{dNi}
\end{eqnarray}
where, explicitly, 
\begin{equation}
p\cdot u=m_{\perp }{\rm cosh} \alpha _{\parallel } {\rm cosh}  \alpha
_{\perp }-p_{\perp }\cos \xi \, {\rm sinh}  \alpha _{\perp }. \label{pu}
\end{equation}

\begin{figure}[t]
\centerline{\includegraphics[width=7cm]{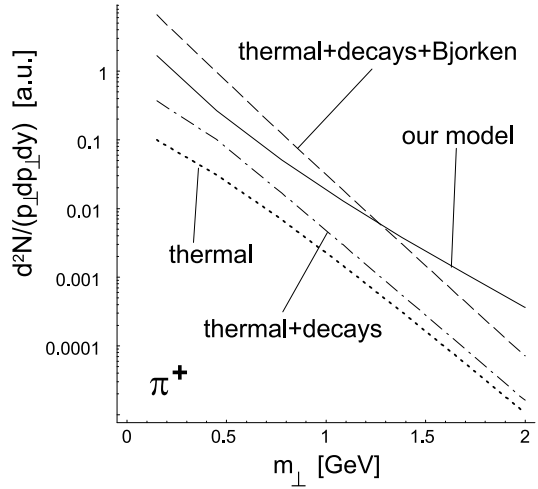}}
\caption{Contributions of various effects to the $m_\perp$-spectra of
positive pions (normalizations arbitrary, the relative norm of dotted
and dash-dotted curves preserved).}
\label{fig:ped}
\end{figure}

We end this section with a pedagogical discussion of the role played
by various effects included.  Figure \ref{fig:ped} shows the $m_\perp$-%
spectrum of positive pions obtained with thermal parameters of Table~%
1. The dotted line shows the spectrum of primordial pions without
expansion.  The dot-dashed line adds the resonance decays; they
contribute about 75\% of the total, with the low momenta more
populated. The dashed line is the result of the model with no
transverse flow, {\em i.e.} including only the longitudinal Bjorken
expansion. Finally, the solid line shows the full calculation, with
resonance decays and the longitudinal plus transverse expansion produced
by parameterization (\ref{tau},\ref{umu}). The characteristic convex
shape is acquired as the result of the transverse flow.

\section{Transverse-momentum spectra\label{sec:spectra}}

\begin{figure}[t]
\centerline{\includegraphics[width=12.5cm]{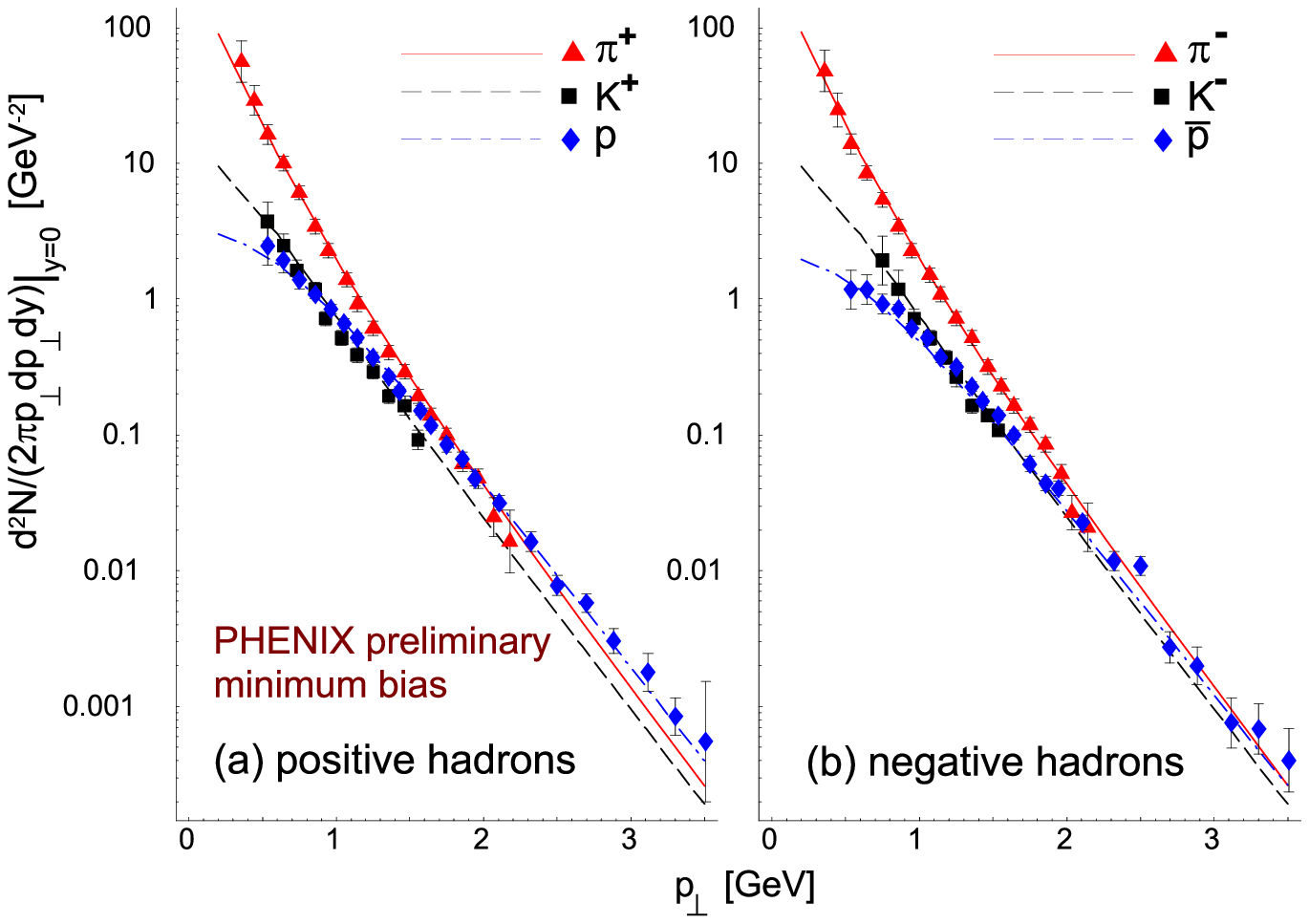}}
\caption{The $p_{\perp}$-spectra of pions (solid line), kaons (dashed line)
and protons or antiprotons (dashed-dotted line), as evaluated from our model, compared to
the PHENIX preliminary data obtained from Fig. 1 of Ref. \cite{phenix}.
Later official PHENIX data of Ref. \cite{phenoff} agree with the data used here.
Feeding from the weak decays is included.}
\label{f0}
\end{figure}
Equipped with all elements of the model, we can now apply it to
describe the $p_\perp$-spectra. The thermal parameters are always
those of Table 1.  In principle, they could change with the centrality
bin (impact parameter), but since the ratios of particles depend
weakly on the centrality \cite{qm01,qm02,hir02}, so do the thermal
parameters. We begin with presenting in Fig. 4 the fit
to the earliest-available minimum-bias data from the PHENIX
collaboration \cite{phenix}.
We observe a very good agreement of our model with the data up to
$p_{\perp }\sim 2$ or even, amusingly, 3~GeV. In that range the model
curves cross virtually all data points within the error bars. At
larger values of $p_{\perp }$, where hard processes are expected to
contribute, the model falls below the data for $p$ and
$\overline{p}$. 
\begin{figure}[b]
\centerline{\includegraphics[width=13.50cm]{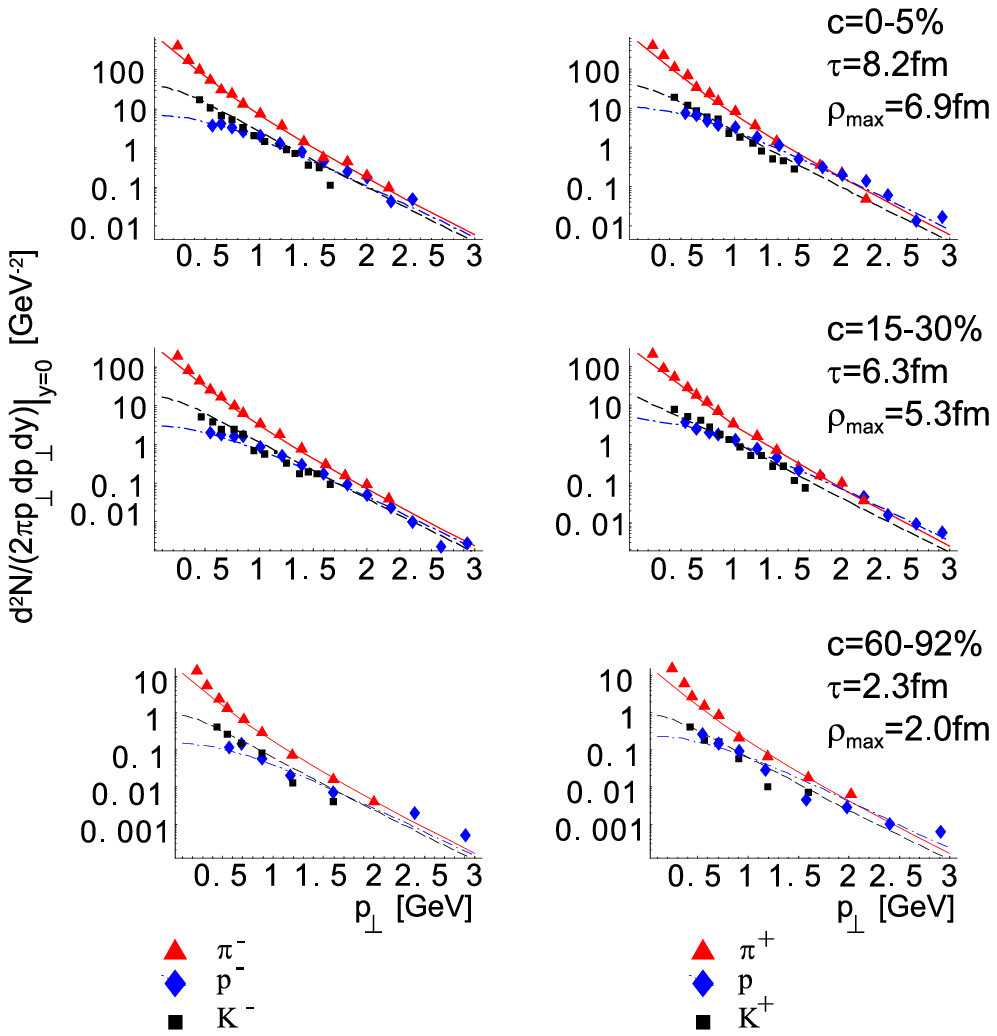}}
\caption{Model vs. experiment for the PHENIX data \cite{phenoff} at three different
centrality bins for pions, kaons, protons and antiprotons. The thermal
parameters are unchanged, while the geometric parameters
following from the fit are given in the figure.}
\label{fcen}
\end{figure}
Since the values of the strange and isospin
chemical potentials are close to zero, the model predictions for $\pi
^{+}$ and $\pi ^{-}$, as well as for $K^{+}$ and $K^{-}$ are virtually the
same. The value of the baryon chemical potential of $41$ MeV splits the $p$
and $\bar{p}$ spectra. Note the convex shape of the pion spectra.
The $\pi^+$ and $p$ curves in Fig. 4 cross at $p_\perp \simeq 2$ GeV, 
and the $K^+$ and $p$ at $p_\perp \simeq 1$ GeV, exactly as in the
experiment. The values of the fitted geometric parameters are
shown in second column of Table 2.

The next plot, Fig. 5, shows an analogous fit made separately for 3 different
centrality bins.  The obtained values of the geometric parameters are
compared in Table 2. Again, the agreement is
satisfactory.\footnote{For non-central collisions the shape of the
hypersurface and the four-velocity at freeze-out is expected to be
deformed in the $x-y$ plane. In fact, in the hydrodynamic approaches
this is the result of the elliptic flow, causing the azimuthal
asymmetry of the spectra. The effect can be incorporated by properly
extending the parameterization (\ref{tau},\ref{umu}). However, the
effect of departing from the cylindrical symmetry by the amount needed
to describe the elliptic-flow coefficient, $v_2$, is negligible for
the $p_\perp$-spectra integrated over the azimuthal angle, considered
in this lecture \cite{abaran}.}

\begin{figure}[b]
\centerline{\includegraphics[width=8.5cm]{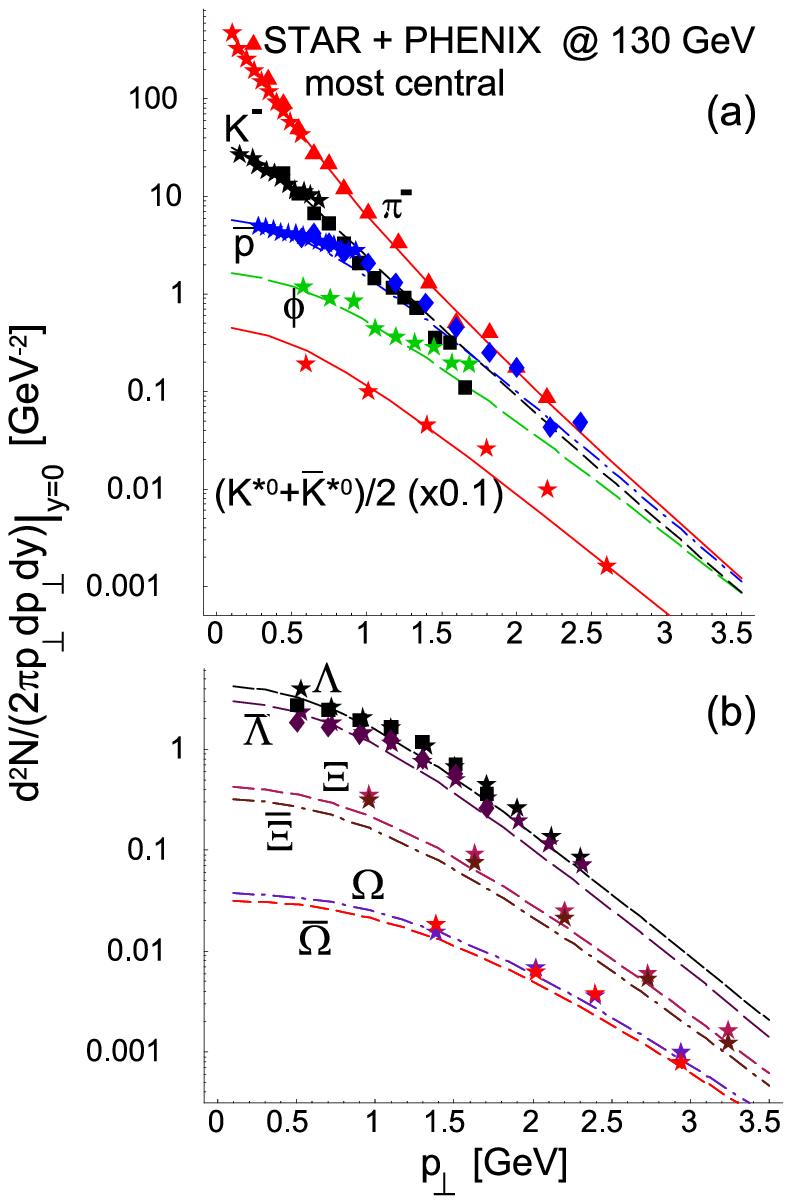}}
\caption{(a) The $p_{\perp}$-spectra at midrapidity 
of $\pi^-$, $K^-$, $\overline{p}$, $\phi$, and $K^*(892)^0$, and (b)
of hyperons $\Lambda$, $\Xi$, and $\Omega$.  
The asterisks represent the data from the STAR, and other symbols from the PHENIX 
collaboration. 
All spectra are for the most central collisions
\cite{phenix,harris,starantip,yama,starLambda,starKstar,phenLambda,castillo,hyppo}.
The STAR data for the $\pi^-$, $K^-$, $\phi$, $\Xi$'s and $\Omega$'s are preliminary. 
The updated experimental points for the $\Xi$'s \cite{Xinew} 
are lower by about a factor of 2 from those shown, 
and much better agreement with the model follows.
All theoretical curves and the data are absolutely normalized. 
The data and the model calculation include full feeding from the weak decays.}
\label{f1}
\end{figure}

Finally, in Fig. \ref{f1} we show our results for all up-to-now
available spectra at $\sqrt{s_{NN}}=130$~GeV for the most-central
collisions, including the particles involving strangeness.  The upper
part of Fig. \ref{f1} displays the spectra of pions, kaons, antiprotons,
used earlier to determine the geometric parameters (last column in
Table 2), and the predicted spectra of the $\phi$ and $K^*(892)^0$
mesons.

The predicted spectrum of the $\phi$ mesons agrees well with the
reported measurement \cite{yama}, with the model curve crossing five out
of the nine data points.  The $\phi$ meson deserves a particular
attention in relativistic heavy-ion collisions, since it serves as a
very good ``thermometer'' of the system. This is because its
interaction with the hadronic environment is negligible. Moreover, it
does not receive any contribution from resonance decays, hence its
spectrum reflects directly the distribution at freeze-out and the
flow. Thus, the agreement of the model and the data for the case of $\phi$ 
supports the idea of one universal freeze-out.

The upper part of Fig. \ref{f1} also shows the averaged spectrum of
$K^*(892)$ resonances, with the data from Ref. \cite{starKstar}.  Once
again we observe a good agreement between the model curve and the
experimental points. As already mentioned in Sec. \ref{sec:model}, the
successful description of both the yield and the spectrum of
$K^*(892)^0$ mesons supports the concept of the thermal description of
hadron production at RHIC, and brings evidence for small interval
between chemical and thermal freeze-outs, in support
of Eq. (\ref{Tuniv}).  If the $K^*(892)^0$ mesons
decayed between the chemical and thermal freeze-out, the emitted pions
and kaons would rescatter and the $K^*(892)^0$ states could not be
seen in the pion-kaon correlations.  In addition, if only a fraction
of the $K^*(892)^0$ yield was reconstructed, it would not agree with
the outcome of the thermal analysis which provides the particle yields
at the chemical freeze-out. Thus, the expansion time between chemical
and thermal freeze-out must be smaller than the $K^*(892)^0$ life-time,
$\tau =$ 4 fm/$c$ \cite{starKstar}.

The bottom part of Fig. \ref{f1} shows the predictions of the model
for the spectra of hyperons. Again, in view of the fact that no extra
parameters have been introduced here and no refitting has been performed,
the agreement is impressive. We note that the preliminary
\cite{castillo} data for the $\Xi$'s used in the figure were
subsequently updated \cite{Xinew}. The following reduction of the data
by about a factor of 2 results in a much better agreement with the
model. The data accumulated at lower energies at SPS showed that the
slope of the $\Omega$ hyperon was much steeper than for other
particles \cite{spsOmega}. On the contrary, in the case of RHIC the
model predictions for the $\Omega$ are as good as for the other
hadrons.  Since the $\Omega$ contains three strange quarks, it is most
sensitive for modifications of the simple thermal model used here,
{\em e.g.} the use of canonical instead of the grand-canonical
ensemble. The agreement of Fig. \ref{f1} does not support the need for
inclusion of these effects.

The various values of the geometric parameter sets are collected in
Table~2. We also show their ratio, as well as the maximum and average
transverse-flow parameter, $\beta$, given in our model by the
equations
\begin{equation}
\beta_\perp^{\rm max}= \frac{\rho_{\rm max}}{\sqrt{\tau^2+\rho_{\rm max}^2}} \label{beta}
\end{equation}
and
\begin{equation}
\langle \beta_\perp \rangle = \frac{\int_0^{\rho_{\rm max}} \rho d\rho 
\frac{\rho}{\sqrt{\tau^2+\rho^2}}}{\int_0^{\rho_{\rm max}} \rho d\rho} . \label{betaav}
\end{equation}
We note that the ratio $\rho_{\rm max}/\tau$, and consequently, $\beta_\perp^{\rm max}$ and
$ \langle \beta_\perp \rangle$,  practically do not depend on centrality. 
\begin{table}[t]
\begin{center}
\begin{tabular}{|r|rrrr|r|}
\hline
& \multicolumn{4}{|c|}{PHENIX} & PHENIX  \\
& \multicolumn{4}{|c|}{} & + STAR  \\
\hline
$c$ [\%]   & min. bias & 0-5 & 15-30 & 60-92 & 0-5/0-6 \\
\hline
\hline
$\tau$ [fm]           & 5.6 & 8.2 & 6.3 & 2.3 & 7.7 \\
$\rho_{\rm max}$ [fm] & 4.5 & 6.9 & 5.3 & 2.0 & 6.7 \\
\hline
$\rho_{\rm max}/\tau$ &   0.81   & 0.84 & 0.84  & 0.87 & 0.87  \\
$\beta_\perp^{\rm max}$ &  0.62  &0.64 & 0.64 & 0.66 & 0.66 \\
$ \langle \beta_\perp \rangle$  & 0.46 & 0.47 & 0.47 & 0.48 & 0.48 \\
\hline
\hline
\end{tabular}
\end{center}
\caption{The fitted values of the geometric parameters for various 
centrality bins, their ratio, and the maximum and average transverse flow 
parameters, as given by Eqs. (\ref{beta},\ref{betaav}).}
\end{table}

To summarize this section, we conclude that the successful and
economic description of the spectra, as seen from
Figs. \ref{f0}, \ref{fcen}, and \ref{f1}, provides a strong support for
the thermal approach with universal freeze-out 
in the description of the ultra-relativistic heavy-ion collisions at RHIC.

\section{Excluded-volume effects\label{sec:exclude}}

In the 
presented model the fitted values for the geometric parameters, 
$\tau$ and $\rho_{\rm max}$, are low, of the order of the size of the colliding nuclei. This
leads to two problems: 1) the values of the HBT radii, as shown in Sec. \ref{sec:HBT} would be
too small compared to the experiment, and 2) there would be little time left for
the system to develop large transverse flow. Both problems can be
solved with the inclusion of the excluded-volume (van der Waals)
corrections.  Such effects were realized to be important already in the
previous studies of the particle multiplicities in ultra-relativistic
heavy-ion collisions \cite{braunmu,vdw,gore}, where they led to a
significant dilution of system.  In the case of the classical
Boltzmann statistics, which is a very good approximation for our
system \cite{mm}, the excluded volume corrections bring in a factor \cite{gore}
\begin{equation}
\frac{e^{-P v_i /T}}{1+\sum_j v_j e^{-P v_j /T} n_j},
\label{vdw}
\end{equation}
into the phase-space integrals, 
where $P$ denotes the pressure, $v_i=4 \frac{4}{3}\pi r_i^3$ 
is the excluded volume for the 
particle of species $i$,
\footnote{The excluded volume per {\em pair} of particles is $\frac{4}{3}\pi
(2r_i)^3$, hence the factor of 4 in the definition of $v_i$.} and
$n_i$ is the density of particles of species $i$.  The pressure must
be calculated self-consistently from the equation
\begin{equation}
P=\sum_i P^0_i(T, \mu_i-P v_i/T)=\sum_i P^0_i(T, \mu_i)e^{-P v_i/T},
\label{pres}
\end{equation}
where $P^0_i$ is the partial pressure of the ideal gas of hadrons 
of species $i$.
For the simplest case where the excluded volumes for all particles are equal, 
$r_i=r$, $v_i =v$, the excluded-volume correction (\ref{vdw}) 
produces a scale factor common to all particles,
which we can denote by $S^{-3}$. The 
formula (\ref{dNi}) becomes
\begin{eqnarray}
\frac{dN_{i}}{d^{2}p_{\perp }dy} &=&\ \tau ^{3}\int_{-\infty }^{+\infty
}d\alpha _{\parallel }\int_{0}^{\rho _{\max }/\tau }{\rm sinh}  \alpha _{\perp
}d\left( {\rm sinh}  \alpha _{\perp }\right) 
\int_{0}^{2\pi }
d\xi \, p\cdot u \, S^{-3} f_{i}\left( p\cdot u\right) .\nonumber \\
\label{dNi2}
\end{eqnarray}
The presence of the factor $S^{-3}$ in Eq. (\ref{dNi2})
may be compensated by rescaling $\rho$ and $\tau$
by the factor $S$. That way, we retain all the previously obtained 
results for the particle abundances and the momentum spectra.
However, now the system is more dilute and larger in size.

Next, we present an estimate of $S$.
With our values of the thermodynamic parameters we have $\sum_i P^0_i(T,
\mu_i)=80$MeV/fm$^{3}$, which leads to $S=1.3$ with $r=0.6$fm. 
Values of this order have been
typically obtained in other works. Thus, the excluded-volume 
corrections can increase the
size parameters at freeze-out by about 30\%, 
and in consequence the problems 1) and 2) are alleviated: the geometric parameters
become large enough to be reconciled with expansion, and the HBT correlation radii
can be properly reproduced, see Sec. \ref{sec:HBT}.

\section{HBT radii\label{sec:HBT}}

\begin{figure}[b]
\centerline{\includegraphics[width=7cm]{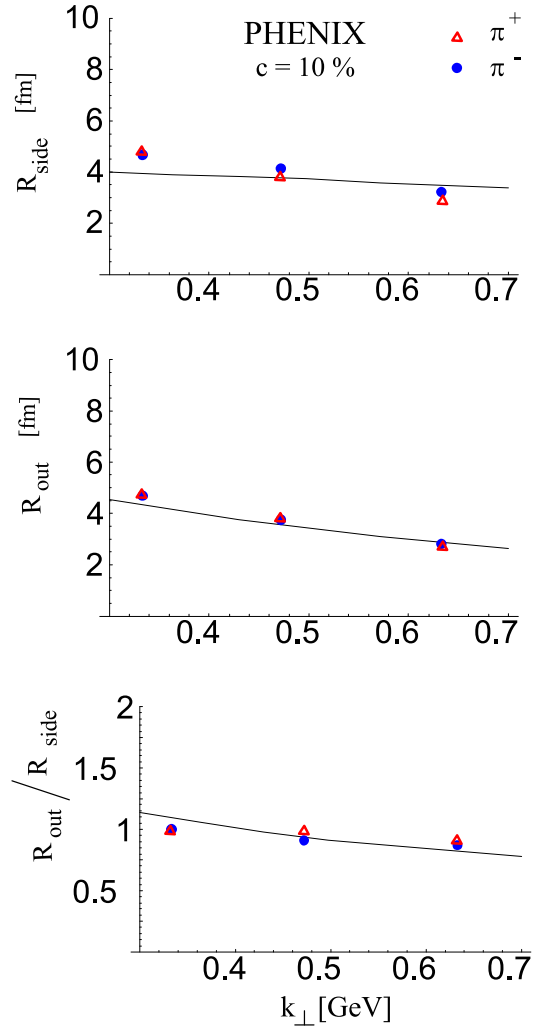}}
\caption{The HBT correlation radii for most-central collisions, 
$R_{\rm side}$, $R_{\rm out}$, and their ratio, as predicted by the model 
(solid line) and measured by the PHENIX collaboration. }
\label{fig:hbt}
\end{figure}

The transverse HBT radii $R_{\rm side}$ and $R_{\rm out}$ (here we use
the Bertsch-Pratt \cite{bp,bp1,be} parameterization) measured
\cite{phenixhbt,starhbt,agshbt} at RHIC have values very close to
those measured at smaller beam energies. Only the longitudinal radius,
$R_{\rm long}$, exhibits a monotonic growth with $\sqrt{s_{NN}}$
\cite{phenixhbt}.  The weak energy dependence of $R_{\rm side}$ and
$R_{\rm out}$ has come as a great puzzle, since the RHIC beam energy,
$\sqrt{s_{NN}}=130$ GeV, is almost one order of magnitude larger than
the SPS energy, $\sqrt{s_{NN}}$ =17 GeV, and based on the
hydrodynamic calculations one would expect much larger systems to be
produced at RHIC. Also, a longer life-time of the
firecylinder was expected at RHIC, which should be reflected in longer
emission times of pions, which in turn would result in the ratio
$R_{\rm out}/R_{\rm side}$ much larger than unity \cite{dumitru}.  On the
contrary, the experimental measurements indicate that
$R_{\rm out}/R_{\rm side}$ is compatible with unity in the whole range of
the studied transverse-momenta of the pion pair ($ 0.2 < k_T < 1.0 $
GeV). This fact is another surprise delivered by the analysis of the
RHIC data for the pion-pion correlations.

We have computed the pion HBT radii in our model. The calculation is
based on the formalism of Ref. \cite{formhbt}, and is similar to the
case of the particle spectra shown in
Sec. \ref{sec:expansion}. Details will be presented elsewhere
\cite{wwhbt}.  The results of an approximate calculation neglecting
the hadronic widths are shown in Fig. \ref{fig:hbt}, where the HBT
radii are plotted as a function of the transverse momentum of the pion
pair, $k_\perp$. We note that very reasonable agreement with the data
is achieved.  We have used $S=1.3$ of Sec. \ref{sec:exclude}.  In
particular, the ratio $R_{\rm out}/R_{\rm side}$
(independent of the scale factor $S$) is close to
unity. The $k_\perp$ dependence of $R_{\rm side}$ is a bit too
flat. The longitudinal radius, $R_{\rm long}$, is sensitive to the
cut-off in the rapidity distribution and cannot be reliably computed
in the present, boost-invariant, model.

\section{Conclusions}

The presented results for the hadron production at RHIC support the
idea that particles are produced thermally, and this is the basic
lesson for today.  The 
simple, economic model with the approximation of a 
universal freeze-out, simple expansion, and complete treatment of resonances,
predicts the particle ratios, the transverse-momentum
spectra, and the HBT correlation radii for the pion in agreement
with the data.  We note that the thermal approach works noticeably
better at the RHIC energies than at lower energies, where, {\em e.g.}, the
particle ratios are not described to that accuracy  \cite{mm}, or the spectrum of
the $\Omega$ baryons is not reproduced. This indicates that the soft
physics becomes simpler at RHIC, with our model being able to yield
the quite impressive results of 
Table 1 and Figs. \ref{f0}, \ref{fcen}, \ref{f1}, and
\ref{fig:hbt}.  

In our phenomenology the pre-freeze-out stages are hidden and only the conditions at the 
moment where the hadrons decouple are relevant. This provides useful constraints for the more
microscopic approaches. These calculations, describing early stages of the evolution,   
should ultimately provide the freeze-out conditions such as, or similar, 
to the ones used in our study. 

Certainly, the most challenging theoretical question which remains and should be 
addressed in future efforts is {\em why the model
works so nicely}, and what it means for the underlying physics of particle production 
and the mechanism of hadronization. 

\bigskip

We are grateful Professor Andrzej Budzanowski for his encouragement and 
interest in this work, 
to Marek Ga\'zdzicki for numerous helpful discussions, and to
Boris Hippolyte for pointing out the early experimental spectra for
the $\Omega$ baryons.


\begin{thebibliography}{100}

\bibitem{qm01} Proceedings of the 15th Int. Conference on Ultrarelativistic Nucleus-Nucleus 
Collisions (Quark Matter 2001), Stony Brook, New York, 15-20 Jan 2001, 
{\it Nucl. Phys.} {\bf A698} (2002).

\bibitem{qm02} Proceedings of the 16th Int. Conference on Ultrarelativistic Nucleus-Nucleus 
Collisions (Quark Matter 2002), Nantes, France, 18-24 July 2002, 
to be published in {\it Nucl. Phys.} {\bf A}.

\bibitem{hir02} Proceedings of the 30th Int. Workshop 
on Gross Properties of Nuclei and Nuclear Excitation: Ultrarelativistic 
Heavy Ion Collisions, Hirschegg, Austria, 13-19 Jan 2002.

\bibitem{braunmu}  P. Braun-Munzinger, J. Stachel, J. P. Wessels, and N. Xu,
{\it Phys. Lett.} {\bf B344}, 43 (1995); {\it Phys. Lett.} {\bf B365}, 1 (1996).

\bibitem{raf} J. Rafelski, J, Letessier, and A. Tounsi, {\it Acta Phys. Pol.} {\bf B28}, 2841 (1997).

\bibitem{cest}  J. Cleymans, D. Elliott, H. Satz, and R. L. Thews, {\it Z. Phys.} {\bf C74}, 319 (1997).

\bibitem{pbmsps}  P. Braun-Munzinger, I. Heppe, and J. Stachel, {\it Phys. Lett.} {\bf B465}, 15 (1999).

\bibitem{yg}  G. D. Yen and M. I. Gorenstein, {\it Phys. Rev.} {\bf C59}, 2788 (1999).

\bibitem{becatt} F. Becattini, J. Cleymans, A. Keranen, E. Suhonen,
and K. Redlich, {\it Phys. Rev.} {\bf C64}, 024901 (2001).

\bibitem{gazgor0} M. Ga\'zdzicki and M. I. Gorenstein {\it Acta Phys. Pol.} {\bf B30}, 2705 (1999).

\bibitem{gaz} M. Ga\'zdzicki, {\it Nucl. Phys.} {\bf A681}, 153 (2001).

\bibitem{raf0} J. Rafelski, J. Letessier, and G. Torrieri, {\it Phys. Rev.} {\bf C64}, 054907 (2001).

\bibitem{pbmrhic}  P. Braun-Munzinger, D. Magestro, K. Redlich, and J.
Stachel, {\it Phys. Lett.} {\bf B518}, 41 (2001).

\bibitem{wfwbmm} W. Florkowski, W. Broniowski, and M. Michalec,
{\it Acta Phys. Pol.} {\bf B33}, 761 (2002).

\bibitem{mm} M. Michalec, PhD Thesis, nucl-th/0112044.

\bibitem{budzan} {\it Acta Phys. Polon.} {\bf B33}, 33 (2002). 

\bibitem{gorj} M. I. Gorenstein, K. A. Bugaev, and M. Ga\'zdzicki, {\it Phys. Rev. Lett.} 
{\bf 88}, 132301 (2002). 

\bibitem{bps} F. Becattini and G. Pettini, hep-ph/0204340.

\bibitem{bg} F. Becattini, {\it J. Phys.} {\bf G28} 1553 (2002). 

\bibitem{cl} J. Cleymans, hep-ph/0201142.

\bibitem{z} D. Zschiesche, S. Schramm, J. Sch\"affner-Bielich, H. St\"ocker, 
and W. Greiner, nucl-th/0209022. 

\bibitem{prorok} D. Prorok, hep-ph/0209235.

\bibitem{wbwf} W. Broniowski and W. Florkowski,
{\it Phys. Rev. Lett.} {\bf 87}, 272302 (2001).

\bibitem{wfepi} W. Florkowski and W. Broniowski, 
{\it Acta Phys. Pol.} {\bf B33}, 1629 (2002). 

\bibitem{hirsch} W. Broniowski and W. Florkowski, in [3], p. 146, hep-ph/0202059.

\bibitem{str} W. Broniowski and W. Florkowski, {\it Phys. Rev.} {\bf C65}, 064905 (2002). 

\bibitem{rhicsps} W. Broniowski and W. Florkowski, {\it Acta Phys. Pol.} {\bf  B33}, 1935 (2002).

\bibitem{wfqm02} W. Florkowski and W. Broniowski,
in [2], nucl-th/0208061.

\bibitem{koppe} H. Koppe, {\it Zs. f. Naturforschung} {\bf 3a}, 251 (1948); {\it Phys.
Rev.} {\bf 76}, 688 (1949).

\bibitem{fermi} E. Fermi, {\it Progr. Theor. Phys.} {\bf 5}, 570 (1950); {\it Phys. Rev.
} {\bf 81}, 683 (1951).

\bibitem{landau} L. Landau, {\it Izv. Akad. Nauk SSSR, Ser. Fiz. } {\bf 17}, 51 (1953).

\bibitem{hagedorn} R. Hagedorn, {\it Suppl. Nuovo Cim. } {\bf 3}, 147 (1965);
preprint CERN 71-12 (1971), preprint CERN-TH. 7190/94 (1994)
and references therein.

\bibitem{gammas} For a recent discussion of this issue see: J. Rafelski and J. Letessier, to 
appear in the Proceedings of Pan American Advanced Studies Institute on New States of Matter in
Hadronic Interactions (PASI 2002), Campos do Jordao, Brazil, 7-18 Jan 2002, hep-ph/0206145.

\bibitem{vdw}   G. D. Yen,  M. I. Gorenstein, W. Greiner, and S.-N. Yang,
{\it Phys. Rev.} {\bf C56}, 2210 (1997).

\bibitem{rafdan} J. Rafelski and M. Danos, {\it Phys. Lett.} {\bf B97}, 279 (1980).

\bibitem{hamieh} S. Hamieh, K. Redlich and A. Tounsi, {\it Phys. Lett.} {\bf B486}, 61 (2000).

\bibitem{rafcrit} J. Rafelski and J. Letessier, {\it J. Phys.} {\bf G28}, 1819 (2002).

\bibitem{heinzr} U. Heinz, {\it Nucl. Phys.} {\bf A661}, 140c (1999), and references therein.

\bibitem{rafelski} J. Rafelski and J. Letessier, {\it Phys. Rev. Lett.} {\bf 85}, 4695 (2000).

\bibitem{starKstar} P. Fachini, STAR Collaboration, nucl-ex/0203019.

\bibitem{starKstar2} C. Adler et al., STAR Collaboration, nucl-ex/0205015.

\bibitem{PDG}  Particle Data Group, {\it Eur. Phys. J.}  {\bf C15}, 1  (2000).

\bibitem{hirano} T. Hirano, in [3].

\bibitem{myhag} W. Broniowski and W. Florkowski, {\it Phys. Lett.} {\bf B490},
223 (2000).

\bibitem{bled} W. Broniowski, in Proc. of Few-Quark Problems, Bled, 
Slovenia, July 8-15, 2000, eds. B. Golli, M. Rosina, and S. \v Sirca, 
p. 14, hep-ph/0008112.

\bibitem{rafhag} A. Tounsi, J. Letessier, and  J. Rafelski, 
contribution to the  NATO Advanced Study Workshop on Hot Hadronic Matter: Theory and Experiment, 
Divonne-les-Bains, France, 27 Jun - 1 Jul 1994, p. 105. 

\bibitem{phobos} B. B. Back, PHOBOS Collaboration,
{\it Phys. Rev. Lett.} {\bf 87}, 102301 (2001).

\bibitem{bearden} I. G. Bearden, BRAHMS Collaboration, {\it Nucl.  Phys.} {\bf A698}, 667c (2002).

\bibitem{harris} J. Harris, STAR Collaboration, {\it Nucl.  Phys.} {\bf A698}, 64c (2002).

\bibitem{caines} H. Caines, STAR Collaboration, {\it Nucl.  Phys.} {\bf A698}, 112c (2002).

\bibitem{ohnishi} H. Ohnishi, PHENIX Collaboration, {\it Nucl.  Phys.} {\bf A698}, 659c (2002).

\bibitem{zxu} Z. Xu,  STAR Collaboration, {\it Nucl.  Phys.} {\bf A698}, 607c (2002).

\bibitem{starphi} C. Adler et al., STAR Collaboration, {\it Phys. Rev.} {\bf C65},
041901 (2002).

\bibitem{starLambda} C. Adler et al., STAR Collaboration, nucl-ex/0203016.

\bibitem{starantip} C. Adler et al., STAR Collaboration, {\it Phys. Rev. Lett. } 
{\bf 87}, 262302 (2001).

\bibitem{suire} C. Suire, STAR Collaboration, in [2].

\bibitem{castilloxi} J. Castillo, STAR Collaboration, in [2].

\bibitem{Karsch} F. Karsch, {\it Nucl.  Phys.} {\bf A698}, 199c (2002). 

\bibitem{nxu} N. Xu and M. Kaneta,  {\it Nucl.  Phys.} {\bf A698}, 306c (2002). 

\bibitem{centr} W. Broniowski and W. Florkowski, {\it Phys. Rev.} {\bf C65}, 024905 (2002).

\bibitem{bjorken} J. D. Bjorken, {\it Phys. Rev.} {\bf D27}, 140 (1983).

\bibitem{baym} G. Baym, B. Friman, J.-P. Blaizot, M. Soyeur, and W.~Czy\.z,
{\it Nucl. Phys.} {\bf A407}, 541 (1983). 

\bibitem{Kolya} P. Milyutin and N. N. Nikolaev, {\it Heavy Ion Phys} {\bf 8}, 
333 (1998); V. Fortov, P.~Milyutin, and N. N. Nikolaev,
{\it JETP Lett.} {\bf 68}, 191 (1998).

\bibitem{siemens} P. J. Siemens and J. Rasmussen, {\it Phys. Rev. Lett.} {\bf 42},
880 (1979); P. J. Siemens and J. I. Kapusta, {\it Phys. Rev. Lett.} {\bf 43},
1486 (1979).

\bibitem{SSH} E. Schnedermann, J. Sollfrank, and U. Heinz,
{\it Phys. Rev.} {\bf C48}, 2462 (1993).

\bibitem{BL} T. Cs\"{o}rg\H{o} and B. L\"{o}rstad, {\it Phys. Rev. } 
{\bf C54}, 1390 (1996).

\bibitem{Rischke} D. H. Rischke and M. Gyulassy, {\it Nucl. Phys.}
 {\bf A697}, 701 (1996); {\it Nucl. Phys.} {\bf A608}, 479 (1996).

\bibitem{SH} R. Scheibl and U. Heinz, {\it Phys. Rev. } {\bf C59}, 1585
(1999).

\bibitem{bugaev} K. A. Bugaev, {\it Nucl. Phys.} {\bf A606}, 559 (1996).

\bibitem{csernai} L. P. Csernai, Zs. I. L\'az\'ar, and D. Moln\'ar,
{\it Heavy Ion Phys.} {\bf 5}, 467 (1997).

\bibitem{neymann} J. J. Neymann, B. Lavrenchuk, and G. Fai,
{\it Heavy Ion Phys.} {\bf 5}, 27 (1997).

\bibitem{magas} V. K. Magas et al., {\it Nucl. Phys.} {\bf A661}, 596c (1999).

\bibitem{CF2} F. Cooper, G. Frye, and E. Schonberg, {\it Phys. Rev.} {\bf D11}, 
192 (1975).

\bibitem{biro} T. S. Bir\'o, {\it Phys. Lett.} {\bf B474}, 21 (2000);
{\it Phys. Lett.} {\bf B487}, 133 (2000).

\bibitem{finland} K.~J.~Eskola, H.~Niemi, P.~V.~Ruuskanen, and
S.~S.~R\"as\"anen, hep-ph/0206230; P.~V.~Ruuskanen, in [2].

\bibitem{csorgohyd} T. Cs\"{o}rg\H{o}, F. Grassi, Y. Hama, and T. Kodama, hep-ph/0204300.

\bibitem{csster} T. Cs\"{o}rg\H{o} and A. Ster, nucl-th/0207016. 


\bibitem{ornik} J. Bolz, U. Ornik, M. Pl\"umer, B.R. Schlei, and
R.M.~Weiner, {\it Phys. Rev.} {\bf D47}, 3860 (1993).

\bibitem{weinhold} W. Weinhold, {\it Zur Thermodynamik des $\pi
N$-Systems}, Diplomarbeit, GSI, Sept. 1995.

\bibitem{SKH} J. Sollfrank, P. Koch, and U. Heinz, {\it Phys. Lett. }{\bf
B252}, 256 (1990).

\bibitem{CF1} F. Cooper and G. Frye, {\it Phys. Rev.} {\bf D10}, 186 (1974).

\bibitem{phenix} J. Velkovska, PHENIX Collaboration, Nucl. Phys. {\bf
A698}, 507c (2002).

\bibitem{phenoff}  K. Adcox et al., PHENIX Collaboration, {\it Phys. Rev. Lett.}
{\bf 88}, 242301 (2002).

\bibitem{abaran} A. Baran, to be published.

\bibitem{yama} C. Adler et al., STAR Collaboration, {\it Phys. Rev.} {\bf C65},
041901 (2002).

\bibitem{phenLambda} K. Adcox et al., PHENIX Collaboration, Phys. Rev. Lett.
{\bf 89}, 092302 (2002).

\bibitem{castillo} J. Castillo, STAR Collaboration, {\it J. Phys. } {\bf G28},
1987 (2002).

\bibitem{hyppo} B. Hippolyte, STAR Collaboration, talk presented at
4th Catania Relativistic Ion Studies, {\it Exotic Clustering},
Catania, Italy, June 10-14, 2002.

\bibitem{Xinew} J. Castillo, STAR Collaboration, in [2].

\bibitem{spsOmega} F. Antinori et al., WA97 Collaboration, {\it
J. Phys. }{\bf G27}, 375 (2001).

\bibitem{gore}  G. D. Yen and M. I. Gorenstein, Phys. Rev. {\bf C59}, 2788
(1999).

\bibitem{bp} S. Pratt, {\it Phys. Rev. Lett.} {\bf 53}, 1219 (1984).

\bibitem{bp1} S. Pratt, {\it Phys. Rev.} {\bf D33}, 72 (1986).

\bibitem{be}  G. Bertsch, M. Gong, and
M. Tohyama, {\it Phys. Rev.} {\bf C37}, 1896 (1988).

\bibitem{phenixhbt} K. Adcox et al., PHENIX Collaboration, {\it Phys. Rev.
Lett. } {\bf 88}, 192302 (2002).

\bibitem{starhbt} C. Adler et al., STAR Collaboration, {\it Phys. Rev. Lett.}
{\bf 87}, 082301 (2001).

\bibitem{agshbt} L. Ahle et al., E-802 Collaboration, nucl-ex/0204001. 

\bibitem{dumitru} S. Soff, S.~A.~Bass, and A. Dumitru, {\it Phys. Rev. Lett.}
{\bf 86}, 3981 (2001).

\bibitem{formhbt} T. Cs\"{o}rg\H{o}, {\it Heavy Ion Phys.} {\bf 15}, 1 (2002).

\bibitem{wwhbt} W. Florkowski and W. Broniowski, to be published.

\end{thebibliography}
\end{document}